\documentclass[preprint,review,12pt]{elsarticle}

\usepackage{graphicx}

\usepackage{amssymb}

\usepackage{lineno}

\usepackage{subfig}

\usepackage{url}

\usepackage{amsmath}

\biboptions{sort&compress}

\journal{Physics Letters B}

\begin{document}

\newcommand{\antip}{\bar{\mathrm{p}}}
\newcommand{\meanpt}{\langle p_T \rangle}
\newcommand{\ptmin}{p_{T \ \mathrm{min}}}
\newcommand{\dndeta}{\frac{dN_{ch}}{d\eta}}
\newcommand{\pt}{p_T}
\newcommand{\ncharged}{N_{ch}}
\newcommand{\kt}{k_T}
\newcommand{\rside}{\mathrm{R}_{\mathrm{side}}}
\newcommand{\rinv}{\mathrm{R}_{\mathrm{inv}}}
\newcommand{\sigmaptcube}{\sigma_S/\meanpt^3}
\newcommand{\csquare}{\mathrm{c}_{\mathrm{s}}^2}
\newcommand{\meankt}{\langle k_T \rangle}
\newcommand{\sigmatcube}{\sigma/T^3}
\newcommand{\epsilontfourth}{\epsilon/T^4}

\begin{frontmatter}


\title{Experimental equation of state in pp and p$\antip$ collisions and phase
transition to quark gluon plasma}


\author[df_unibo,infn_bo]{Renato Campanini\corref{cor1}}
\ead{Renato.Campanini@bo.infn.it}

\author[df_unibo]{Gianluca Ferri}
\ead{g.ferri@unibo.it}

\cortext[cor1]{Corresponding Author: Tel. +39 051 2095078,
Fax +39 051 2095047,\\
Mobile +39 3485925020}

\address[df_unibo]{Universit\`{a} di Bologna, Dipartimento di Fisica, viale C. Berti Pichat 6/2, I-40127, Bologna, Italy}
\address[infn_bo]{INFN, Sezione di Bologna, viale C. Berti Pichat 6/2, I-40127, Bologna, Italy}

\begin{abstract}
We deduce approximate equations of state from experimental measurements in
pp and p$\antip$ collisions.
Thermodynamic quantities are estimated combining the measure of average
transverse momentum
$\meanpt$ vs pseudorapidity density $\dndeta$ with the estimation
of the interaction region size from measures of Bose Einstein correlation,
or from a
theoretical model which relates $\dndeta$ to the impact parameter.
The results are very similar
to theory predictions in case of crossover from hadron gas to quark gluon
plasma. According to our analysis, the possible crossover should start at
$\dndeta \simeq 6$ and end at $\dndeta \simeq 24$.
\end{abstract}


\end{frontmatter}

\linenumbers

\section{Introduction}
\label{sec:intro}
Some of the most important questions about the transition to the quark gluon
plasma (QGP), a new state of matter with partonic degrees freedom, are not  yet
fully answered. Among them the location of phase boundaries between hadronic gas
and the QGP. The results of lattice QCD simulations concerning the order of
phase transition depend strongly on the number of quark flavors and on the quark
masses. For vanishing baryon chemical potential $\mu_b = 0$, the nature of
transition can be a genuine phase transition (first order or continuous), or
just a rapid change (crossover) over a small temperature
range~\cite{Aoki2006a}. Estimates of energy densities which can be achieved in
ultra-relativistic pp or p$\bar{\mathrm{p}}$ collisions with high multiplicities
suggest values sufficiently high for experimental formation of the
QGP~\cite{Redlich1986}.

However it may be that, unlike what happens in heavy ion interactions, in pp and
p$\antip$ the central blob of created matter never
thermalizes~\cite{Braun-Munzinger2007},
although there are different
opinions~\cite{Redlich1986, VanHove_2_1983, Werner_ridge_pp_2011,
Bjorken_formula_1983, Castorina_thermalization_2007} which
predict that thermodynamics concepts may be applied in pp or p$\antip$
high multiplicity events.

\begin{figure}
	\begin{center}
	\subfloat[]{
		\label{fig:theoretical_sigmapt3}
		\includegraphics[width=0.35\textwidth]{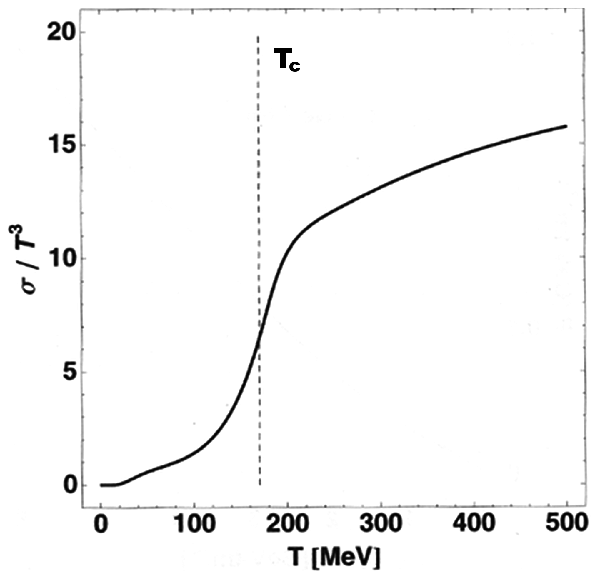}
	}   
	\subfloat[]{
		\label{fig:theoretical_c2s1}
		\includegraphics[width=0.35\textwidth]{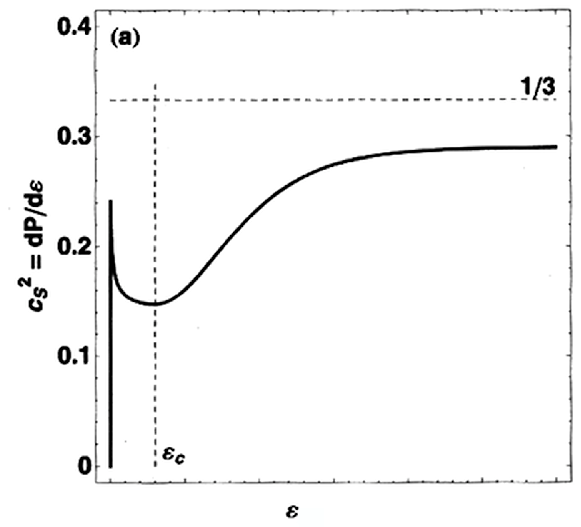}
	}\\
	\subfloat[]{
		\label{fig:theoretical_c2s2}
		\includegraphics[width=0.65\textwidth]{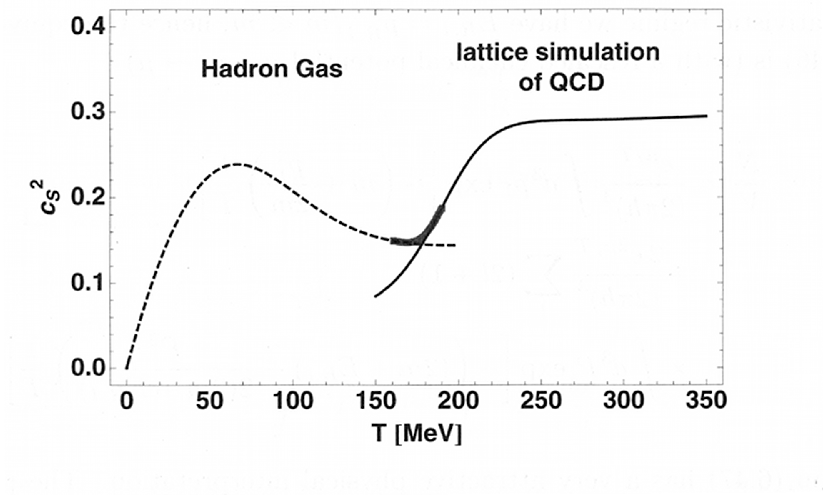}
	}
	
	\caption{ Results of the lattice simulations of QCD for $T > T_c$ (critical
	temperature) and from ideal hadron-gas model for $T < T_c$.\\
	\ref{fig:theoretical_sigmapt3}: entropy density $\sigma$ scaled by $T^3$
calculated in the hadron-gas
model and by lattice simulations of QCD shown as function of temperature.
The vertical line indicates the critical temperature.\\
\ref{fig:theoretical_c2s1}: sound velocity $\csquare$ shown as function of the energy
density $\epsilon$.\\
\ref{fig:theoretical_c2s2}: temperature dependence of the square of the sound
velocity at zero baryon density as function of $T$.
In this case the critical temperature $T$ is equal to 170 MeV.\\
From~\cite{Florkowski_BOOK_2010}.}
	\label{fig:theoretical_lattice}
	\end{center}
\end{figure}

Probes of equation of state are among
possible signatures of phase transition or crossover. The basic idea behind this
class of signatures is the identification of modifications in the dependence of
energy density $\epsilon$, pressure $P$ and entropy density $\sigma$ of hadronic
matter
on temperature $T$. One wants to search for a rapid rise in the effective number
of degrees of freedom, as expressed by the ratio $\epsilontfourth$ or
$\sigmatcube$, over a small temperature range. One can expect a step-like
rise
as predicted by lattice simulations (Fig.~\ref{fig:theoretical_lattice}), more
or less steep depending from the presence of transition or crossover, and
from the order of the transition in the former case.
Finite volume effects may cause important consequences for 
$\epsilontfourth$
and $\sigmatcube$: the latent heat and the jump in the entropy density
are considerably reduced for small systems~\cite{Elze1986}.
Besides that, the critical
temperature may shift to higher temperatures and the width of the
transition may broaden for smaller volumes and there may be a
smoothening of singularities due to the finite size of the
system~\cite{Elze1986,Bazavov2007,Palhares2010}.

In 1982 it has been suggested by~\citet{VanHove_1_1982} that
an anomalous behavior of average transverse momentum
$\meanpt$ as function of the multiplicity could be a
signal for the occurrence of a phase transition in hadronic matter. His
conjecture is based on the idea that the $\meanpt$ distribution of secondaries
reflects the temperature of the system and its evolution in the transverse
direction, while the multiplicity per unit rapidity provides a measure of
entropy~\cite{Muller1995,Mohanty2003}. In  a recent
paper~\cite{Campanini_1_2010} one of us showed that from 22 to 7000~GeV in  21
$\meanpt$ vs pseudorapidity density $\dndeta$ curves there is a
slope change at $\dndeta = 5.5 \pm 1.2$.
Signals related to these slope changes may indicate transition to a new
mechanism of particle production.
Many years ago, in~\cite{Campanini_2_1985}, we pointed out that pp
at ISR and p$\antip$ data at CERN collider showed a kind of jump at $\dndeta =
6$
and that it had to be investigated as a possible phase transition
signal~\cite{Shuryak_Softest_Point_1986}.
In 2002, \citet{Alexopoulos2002} assumed that the
system produced in p$\antip$ at $\sqrt{s} = 1800$ GeV for $\dndeta > 6.75$ was
above the deconfinement transition to explain their experimental results.

In present article, taking into account experimental results in pp and
p$\antip$ at high
energies~\cite{ALICE_PTVSN_Aamodt2010, CMS_PTVSN_Khachatryan2011,
ATLAS_PTVSN_1_Aad2010, ATLAS_PTVSN_2_Aad2011, ALICE_BE_1_Aamodt2010,
ALICE_BE_2_2011,
CMS_BE_1_Khachatryan2010,
CMS_BE_2_2011, CDF_PTVSN_RUN1_Acosta2002,
CDF_PTVSN_RUN2_Aaltonen2009a, E735_PTVSN_Alexopoulos1993, UA1_PTVSN_Albajar1990,
ISR_SFM_PTVSN_Breakstone1987,STAR_pp_BE_2010}, we show how measured physical
quantities satisfy
relations which, given proper approximations and correspondences, can give a
representation of the equations of state (EOS) that describe the created system
in the central region in pseudorapidity in high energy pp and p$\antip$
collisions.
Starting from $\meanpt$~vs~$\dndeta$ experimental results together with the
estimation of the size $S$ of the
interaction area, which is obtained from the measurements of the radii of
emission in function of
multiplicities~\cite{ALICE_BE_1_Aamodt2010, ALICE_BE_2_2011,
CMS_BE_1_Khachatryan2010,CMS_BE_2_2011}, or from a model which relates
multiplicity to impact parameter~\cite{Bialas_impact_parameter}, we
obtain relations among $\meanpt$ and particle density
$\sigma_S$, which
seem to resemble EOS curves predicted for hadronic matter with crossover to QGP.
The $\meanpt$ and $S$ vs~$\dndeta$ relations contain the relevant
information, which translates in $\meanpt$~vs~$\sigma_S$ correlations.

According to our knowledge, this is the first attempt to obtain an estimation of
the complete EOS for hadronic matter using experimental data only.

\section{Methods}
\label{sec:methods}
The experimental results and the approximations made in this work are the
following.

\subsection{$\meanpt$~vs~$\dndeta$}
\label{subsec:methods_ptvsntemperature}
As we mentioned before, $\meanpt$ vs $\dndeta$ correlation at about
$\dndeta = 6$ shows a slope change in all the experiments.
In Van Hove scheme $\meanpt$ reflects temperature and the system evolution. On
the other hand, the biggest part of emitted particles is constituted by pions
and the pion $\meanpt$ is rather insensitive to flow~\cite{Shuryak_book_2003}.
Thus, not identified charged particles $\meanpt$ may be considered as an
estimation of the system temperature because it's not influenced very much
by a possible transverse expansion. Furthermore, transverse
radius $\rside$ vs pair transverse momentum $\kt$ in pp Bose Einstein
correlation measures~\cite{ALICE_BE_2_2011} show that, at least until $\dndeta
\simeq 3.4$, results
are consistent with the absence of transverse expansion, which further supports
the adoption of $\meanpt$ as an identifier of temperature, because it's little
affected by the expansion.

Since a substantial number of pions is the product of resonance decay and the
particles originating from the resonance decays  populate the low $\pt$
region~\cite{Florkowski_BOOK_2010}, in this work we consider
mainly $\meanpt$~vs~$\dndeta$ correlations with a
$\ptmin$ cut ($>$ 400 MeV/c in CDF experiments Run I and Run II , $>$ 500 MeV/c 
in ALICE, and two different cuts, $>$ 500 MeV/c  and $>$ 2500
MeV/c, in ATLAS experiment) in order to work with
$\meanpt$ values less influenced by this effect.
Furthermore, diffractive events are substantially reduced for
$\dndeta \gtrsim 2$ in $\meanpt$ vs $\dndeta$ plots with $\ptmin \geq$
400~MeV/c~\cite{ATLAS_PTVSN_1_Aad2010, ATLAS_PTVSN_2_Aad2011}.

We will show anyway also some results for $\meanpt$ computed with $\ptmin$
cut 0 and $>$ 100 MeV/c. The structure of the relations we are going to show
is still present in these measures.

In this work, $\meanpt$ computed for different $\pt$ cuts will be plotted
without the application of corrections due to the cut in the used $\pt$ range,
apart from the case of events energy density estimation, in which we will use a
corrected $\meanpt$. Regarding $\dndeta$, it is computed from the number of
particle in a given region of pseudorapidity $\eta$ and $\pt$, dividing by the
amplitude of the $\eta$ range
and properly correcting for $\pt$ cuts. In order to perform this last
correction,
we considered $\dndeta$ curves for the different experiments, measured with and
without $\pt$ cuts, and multiplied by the ratio between correspondent values of
$\dndeta$ in the central region.
All data are obtained from minimum bias experiments. For CDF run II 1960 GeV,
high multiplicity trigger data are added to minimum bias data, for
charged particle multiplicity $\ncharged \geq$ 22 ( $|\eta|<1$, $\pt >$ 400 MeV/c
corresponding to $\dndeta$ corrected $\geq$ 22)~\cite{Campanini_1_2010,
CDF_PTVSN_RUN2_Aaltonen2009a}.

Where available, we considered also raw data results (i.e. computed without
experimental inefficiencies corrections)
because the $\meanpt$~vs~$\dndeta$ plot and its derived plots are much
sensible to experimental
losses and, on the other hand, the application of corrections may involve some
``smearing'' of data which could highly modify the analyzed
effects~\cite{ALICE_PTVSN_Aamodt2010,Moggi_thesis_1999}. For reasons of space
we don't show behaviors for raw data in this paper, because results are 
very similar to those for corrected data.

\subsection{Entropy Density Estimation}
\label{subsec:methods_entropy}
The initial energy density in the rest system of a head-on collision has been
argued to be~\cite{Bjorken_formula_1983}:
$$ \epsilon \simeq  \frac{\dndeta \cdot \frac{3}{2} \meanpt}{V} $$
$V$ denotes the volume into which the energy is deposited.
Similarly the initial entropy density is~\cite{Redlich1986}:
$$\sigma \simeq \frac{\dndeta \cdot \frac{3}{2}}{V}$$
As a result, $\epsilon$ is equal to $\sigma \cdot \meanpt$.
The volume $V$ may be estimated as  $V = S \cdot ct$, where $S$ is the
interaction area and $ct$ is a longitudinal dimension we can traditionally
consider to be about 1~fm long.

In order to study our system, we will use the quantity $$\sigma_S =
\frac{\dndeta
\cdot \frac{3}{2}}{S}$$ as an estimation of entropy density. In models like
color glass condensate
and percolation, the system physics depends on
$\sigma_S$~\cite{McLerran1994,DiasDeDeuse_Percolation_Deconfinement_2006,
Satz_Percolation_Deconfinement_2000}.
For the estimation of the area of interaction $S$, we proceed in different ways,
our target being the obtainment of results which are robust respect to the
definition of the area. On the other hand, we are more interested in relations
between variables than in their absolute values.

\subsection{Bose Einstein correlation for emission region size estimation}
\label{subsec:methods_besize}
\begin{figure}
	\begin{center}
	\subfloat[]
	{\includegraphics[width=0.6\textwidth]{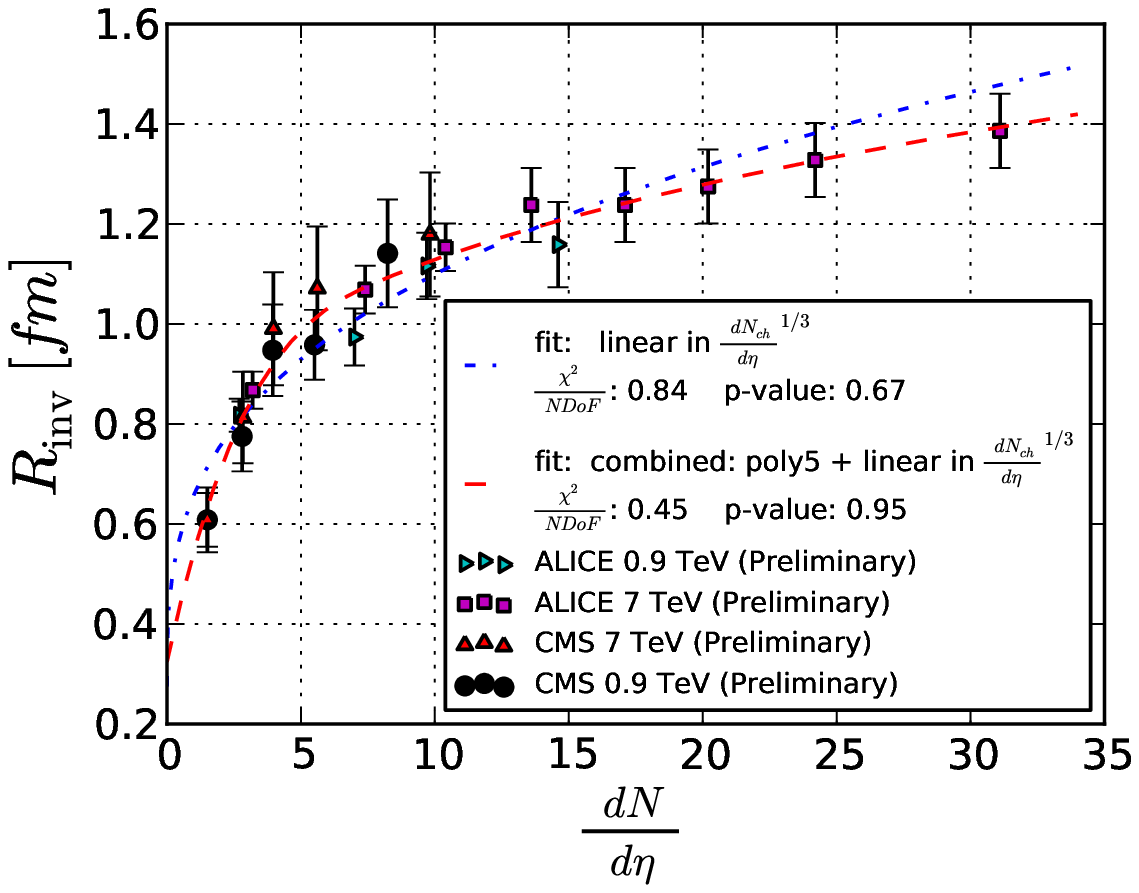}\label{fig:rinv_fit}}\\
	\subfloat[]
	{\includegraphics[width=0.6\textwidth]{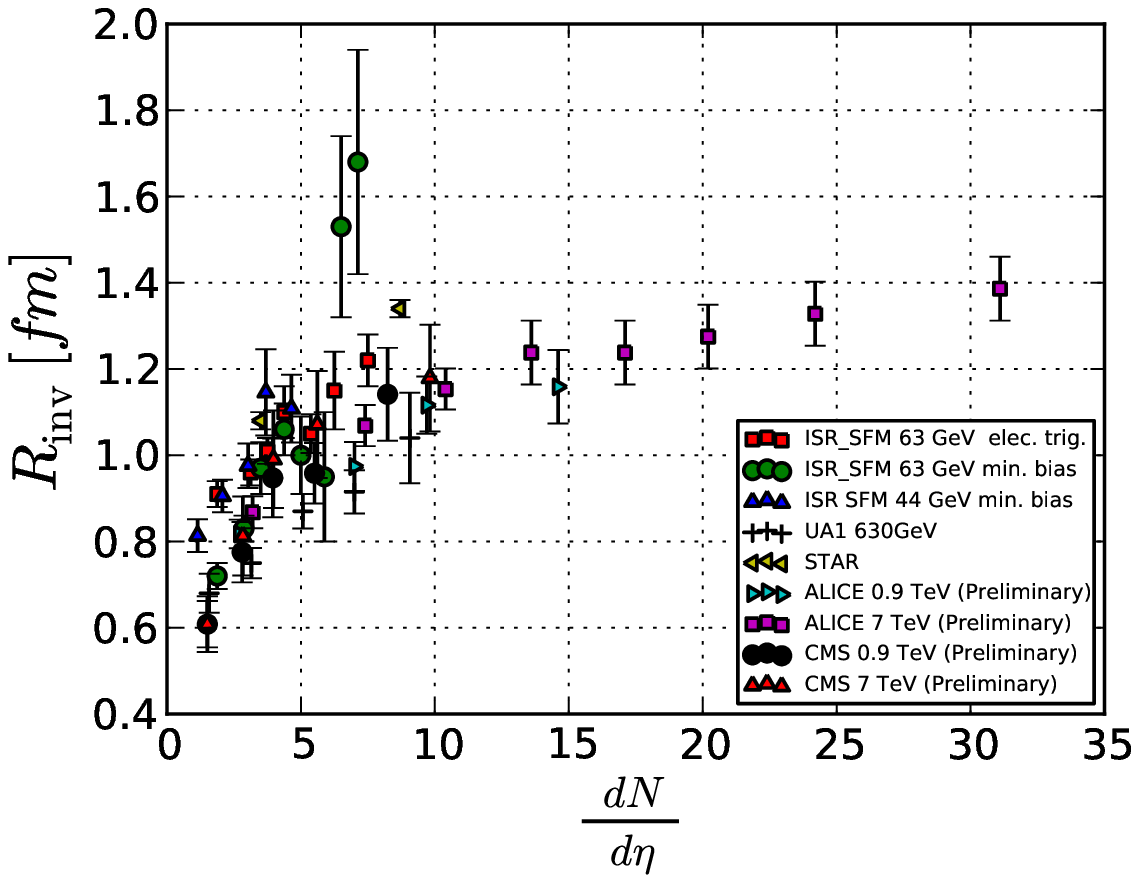}\label{fig:rinv_all}}
	\caption{$\rinv$ vs $\dndeta$ for different
	experiments~\cite{CMS_BE_1_Khachatryan2010,CMS_BE_2_2011,
ALICE_BE_1_Aamodt2010, ALICE_BE_2_2011, ISR_SFM_PTVSN_Breakstone1987,
UA1_BE_Albajar1989,STAR_pp_BE_2010}. Left plot shows data from CMS (preliminary)
and
ALICE(preliminary) both at $\meankt \simeq 0.35$,
along with linear fit in $\dndeta^{\frac{1}{3}}$ and
a combined fit: $5^{\mathrm{th}}$ degree polynomial for $\dndeta < 7.5$ matched
to linear fit in $\dndeta^{\frac{1}{3}}$ for $\dndeta > 7.5$. Right plot shows
data from the first plot along with data from UA1, ABCDHW ISR, and STAR
experiments.}
	\label{fig:rinv}
	\end{center}
\end{figure}

Using Bose Einstein correlation among emitted particles, measurements of
particle emission regions in many pp and p$\antip$ experiments have been
done~\cite{ALICE_BE_1_Aamodt2010,
ALICE_BE_2_2011,CMS_BE_1_Khachatryan2010, CMS_BE_2_2011,  
ISR_SFM_PTVSN_Breakstone1987, STAR_pp_BE_2010, E735_BE_Alexopoulos1993a,
UA1_BE_Albajar1989, BE_RACCOLTA_Chajecki2009}
In ~\cite{ALICE_BE_1_Aamodt2010,ALICE_BE_2_2011}, as already
mentioned, the
measurement of $\rside$ in function of $\kt$ in pp, shows that the transverse
radius doesn't depend on $\kt$ for low $\dndeta$ values ( $<$ 3.4). This can be
explained by
the absence of expansion of the particle emission source, at least at these
$\dndeta$ values.
For $\dndeta$ values greater than 7, there is a dependence of $\rside$ on $\kt$,
so that probably a source expansion is possible at least from this $\dndeta$
value.
It is thus possible that a new phenomenon is started in events with $\dndeta$
between 3.4 and 7.
The hypothesis of no expansion for low $\dndeta$ values lets us
approximate the initial interaction section radius to be coincident to the final
emission radius. We take into account that for $\dndeta$ values greater than
about 7.5, this approximation is more uncertain.
Furthermore, resonance effects are
present, but, once more, we are not interested in the absolute values of the
interaction section, but in its behavior in function of $\dndeta$. Not taking
into account these effects yields a systematic error on the value of the radius,
which we consider invariant for different values of $\dndeta$. Given the
similarity in both the behavior and the absolute values of $\rside$ and
invariant radius $\rinv$ versus multiplicity, we use $\rinv$ as an estimation of
the
interaction region radius, mainly because $\rinv$ data where measured for a
larger $\dndeta$ range than $\rside$ ones~\cite{
 ALICE_BE_1_Aamodt2010, ALICE_BE_2_2011,
CMS_BE_1_Khachatryan2010,
CMS_BE_2_2011,
ISR_SFM_PTVSN_Breakstone1987,  STAR_pp_BE_2010, UA1_BE_Albajar1989,
E735_BE_Alexopoulos1993a}.
In Fig.~\ref{fig:rinv}, $\rinv$ is shown as a function of pseudorapidity
density. In the left we
only show data for CMS (preliminary) and ALICE (preliminary) at
$\meankt \simeq 0.35$, while in the right
side we show the same results along data from other experiments (UA1, ABCDHW
ISR, STAR).
We fitted the data of Fig.~\ref{fig:rinv}a with two functional relations between
$\rinv$ and
$\dndeta$: the first is linear in the cube root of
$\dndeta$~\cite{ALICE_BE_1_Aamodt2010, ALICE_BE_2_2011, Lisa2008} and
the second is linear in cube root of $\dndeta$ for $\dndeta > 7.5$, matched with
a $5^{\mathrm{th}}$ 
degree polynomial fit for smaller $\dndeta$ values. The first fit gives a
$\frac{\chi^2}{NDoF}$ = 0.84 with $p$-value: 0.67, while the second gives a
$\frac{\chi^2}{NDoF}$ = 0.45 with $p$-value: 0.95.
Considering these results, we opted to use the second fit for the following
analysis.

It has been stated that the behavior
of radii in function of $\dndeta$ doesn't depend on the experiment
energy~\cite{BE_RACCOLTA_Chajecki2009}. Data in
Fig.~\ref{fig:rinv} seem to confirm this statement, and justify our choice of a
single relation for $\rinv$~vs~$\dndeta$ for all energies.

In order to estimate the interaction region, we used the following alternatives:
\begin{enumerate}
 \item
 An area obtained using $\rinv$ from the combined (polynomial + linear in
cube root of $\dndeta$) fit from Fig~\ref{fig:rinv_fit}. This choice may
overestimate the interaction region in case of system expansion, being $\rinv$ a
measure of the emission region;
\item
following ALICE results in $\rside$~vs~$\kt$, we make the hypothesis that no
expansion is present in events with sufficiently low $\dndeta$. So we use
$\rinv$ from the left (polynomial) part of the combined fit in
Fig~\ref{fig:rinv}a, then we use
a constant radius for $\dndeta > 7.5$ as an estimation of the dimensions of the
initial region before the possible expansion, making the assumption that at
$\dndeta \simeq 7.5$ the interaction region reaches its maximum;
at $\dndeta = 7.5$ the $\rinv$ value is $1.08$~fm;
\item
an area obtained from a model which relates the impact parameter to the
multiplicity of events~\cite{Bialas_impact_parameter}.

\end{enumerate}

From $\dndeta$ values and from interaction areas, estimated as described above,
we obtained the values of density of particles for transverse area.

We considered the simplified case where the central blob volume $V$ is the same
in all collisions for a given $\dndeta$~\cite{VanHove_1_1982}. We estimate an
average $\sigma$ from the ratio between $\dndeta$ and the estimated
average $V$.

\subsection{$\sigmaptcube$ vs $\meanpt$}
\label{subsec:methods_sigmaptcubevspt}
Using the estimated $\sigma_S$, the relation $\meanpt$~vs~$\sigma_S$ can be
studied. A slope change in $\meanpt$~vs~$\sigma_S$ plots is found at
$\sigma_S$ between
2.5 and 3 $\mathrm{fm}^{-2}$, depending on the method used for the estimation
of area $S$ and corresponds directly to the slope change seen
in $\meanpt$~vs~$\dndeta$ at $\dndeta \simeq 6$.

\begin{figure}
	\begin{center}
	\subfloat[]{
		\label{fig:sigmapt3_alice_09_pt05}	
\includegraphics[width=0.45\textwidth]
{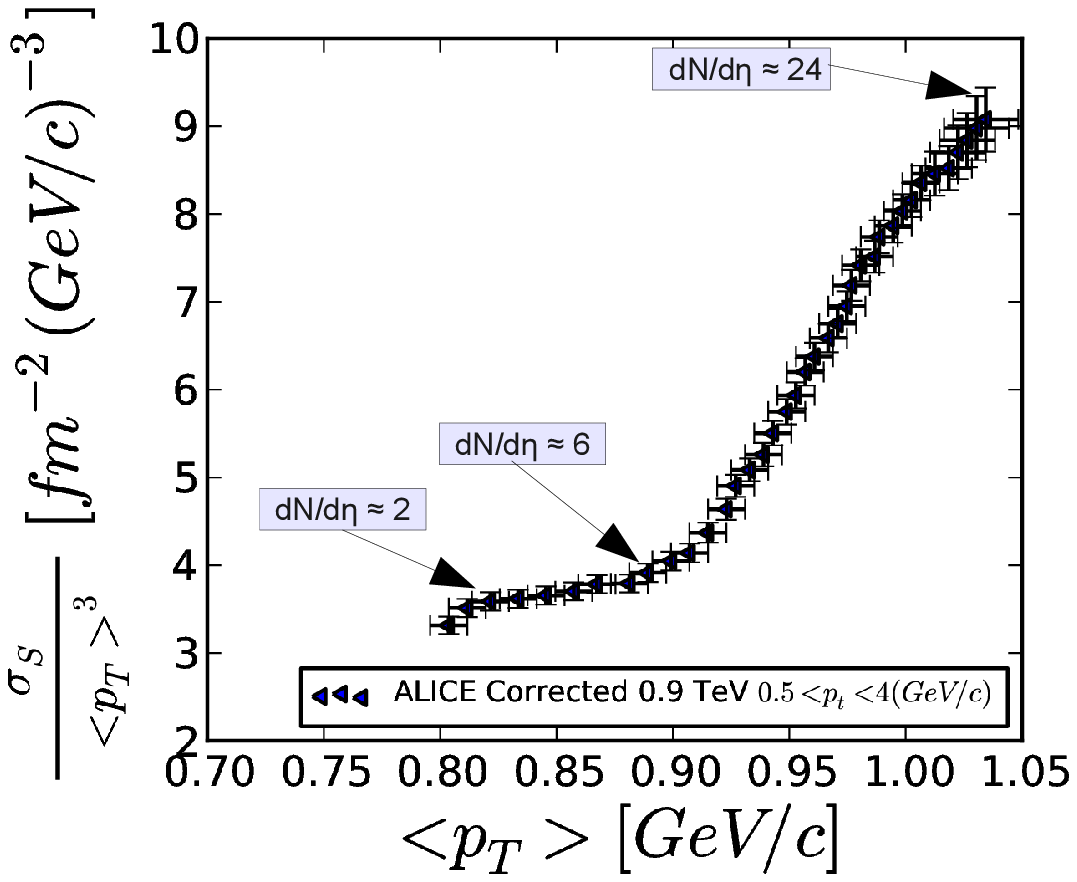}
	} 
\subfloat[]{
		\label{fig:sigmapt3_atlas_09_pt05}
		\includegraphics[width=0.45\textwidth]
		{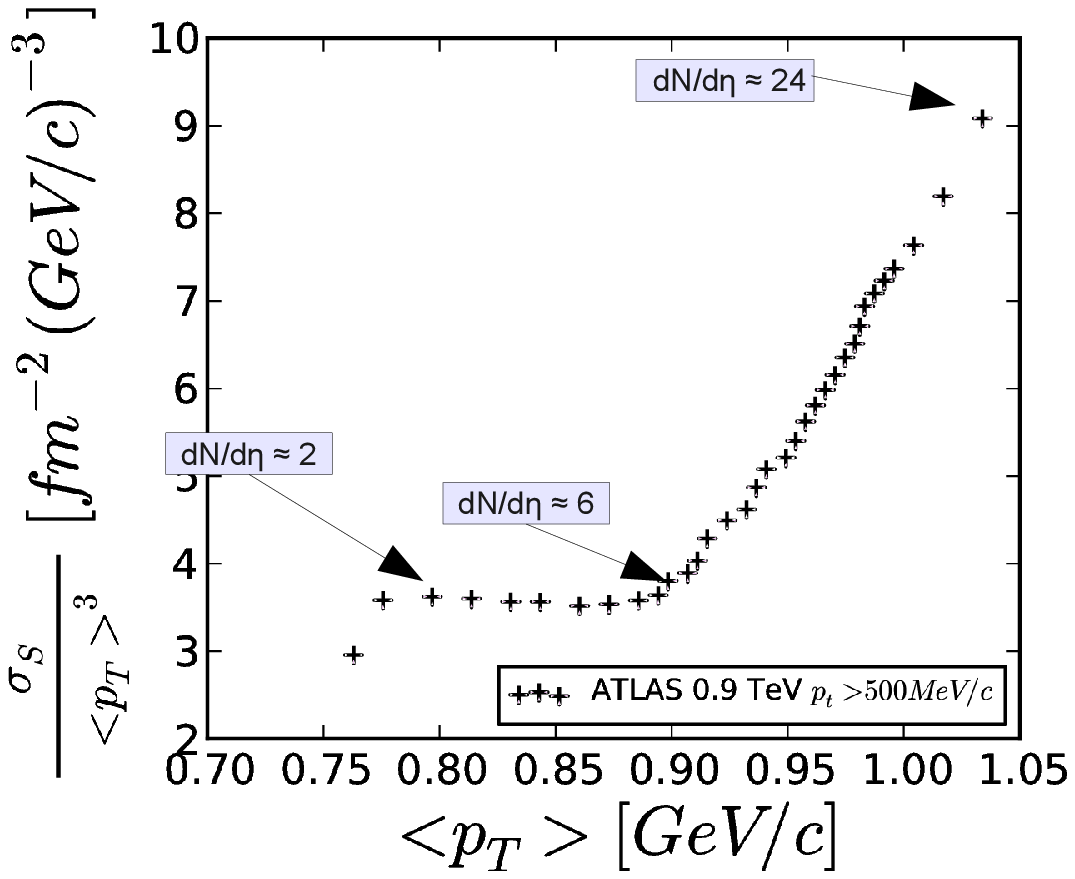}
	}\\
	\subfloat[]{
		\label{fig:sigmapt3_atlas_7_pt05}
	\includegraphics[width=0.45\textwidth]
{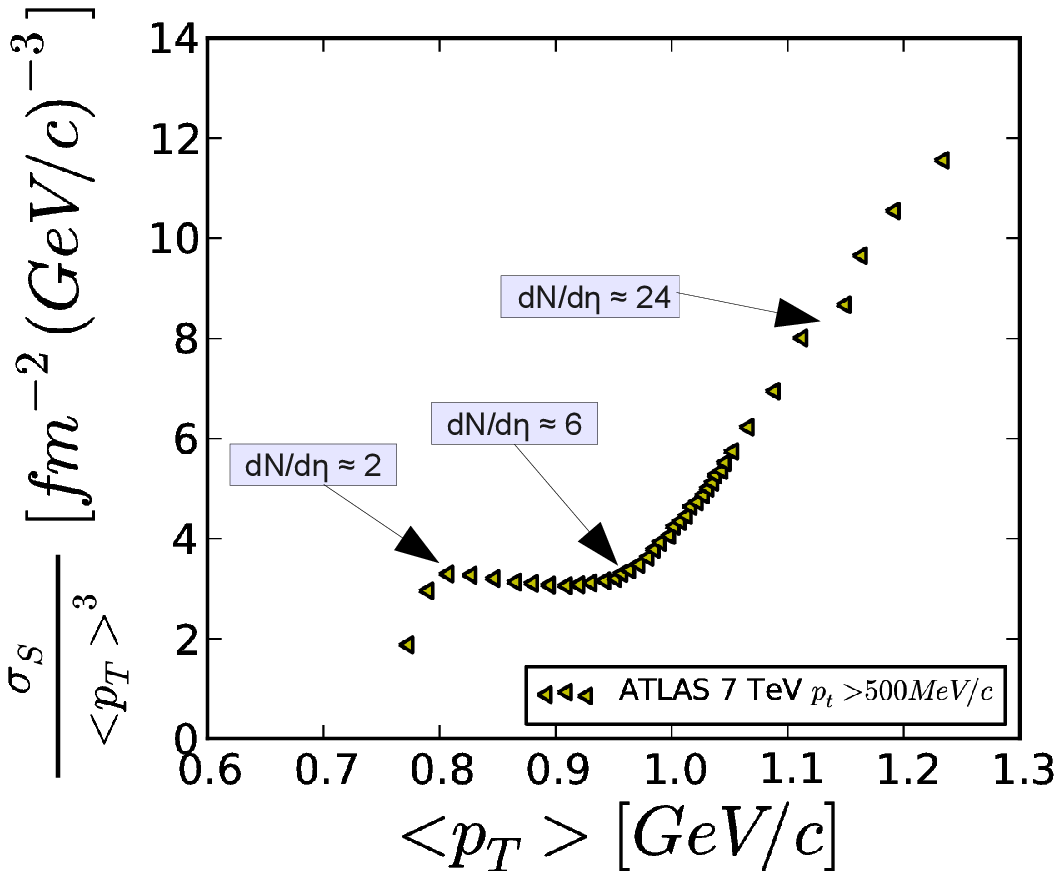}
	}   
	\subfloat[]{
		\label{fig:sigmapt3_atlas_7_pt2500}
	\includegraphics[width=0.45\textwidth]
{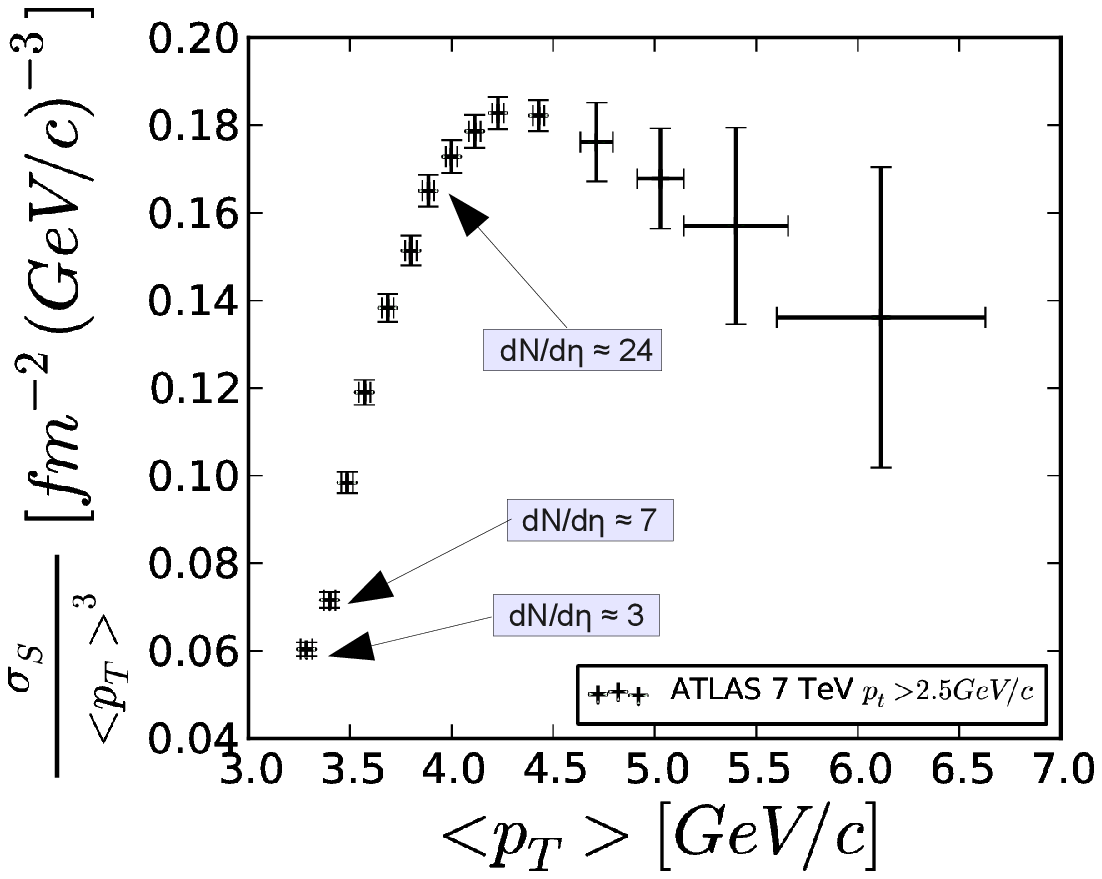}
	}
	
	\caption{$\sigmaptcube$~vs~$\meanpt$ plots. Area $S$ from ``$5^\mathrm{th}$
	degree polynomial + constant after $\dndeta > 7.5$'' fit.\\
	\ref{fig:sigmapt3_alice_09_pt05}: ALICE at $\sqrt{s}$ = 0.9 TeV, $0.5 < \pt < 4$ GeV/c, $|\eta| < 0.8$,
	Minimum Bias.\\
	\ref{fig:sigmapt3_atlas_09_pt05}: ATLAS at $\sqrt{s}$ = 0.9 TeV, $\pt > 0.5$ GeV/c, $|\eta| < 2.5$,
	Minimum Bias.\\
	\ref{fig:sigmapt3_atlas_7_pt05}: ATLAS at $\sqrt{s}$ = 7 TeV, $\pt > 0.5$ GeV/c, $|\eta| < 2.5$,
	Minimum Bias.\\
	\ref{fig:sigmapt3_atlas_7_pt2500}: ATLAS at $\sqrt{s}$ = 7 TeV, $\pt > 2.5$ GeV/c, $|\eta| < 2.5$,
	Minimum Bias.
	}
	\label{fig:sigmaptcube_1}
	\end{center}
\end{figure}

\begin{figure}
	\begin{center}
	\subfloat[]{
		\label{fig:sigmapt3_cdf_run1}	
\includegraphics[width=0.45\textwidth]
{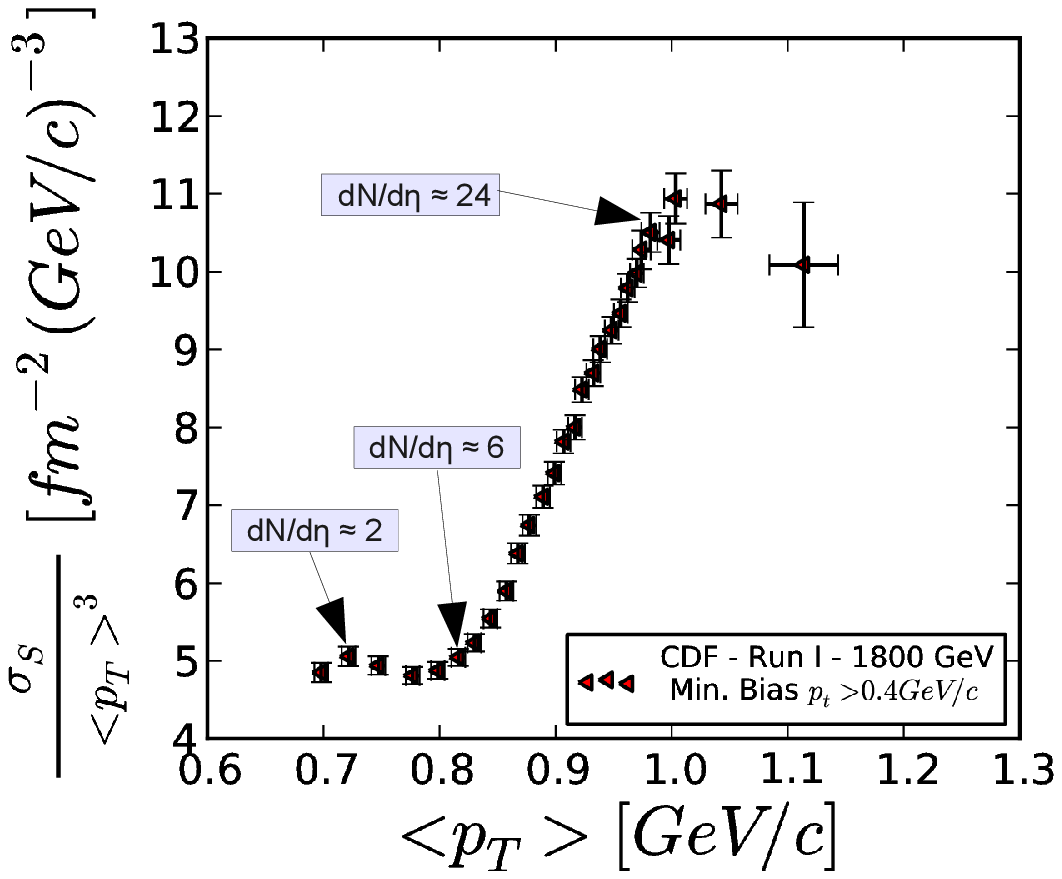}
	} 
\subfloat[]{
		\label{fig:sigmapt3_cdf_run2}
		\includegraphics[width=0.45\textwidth]
		{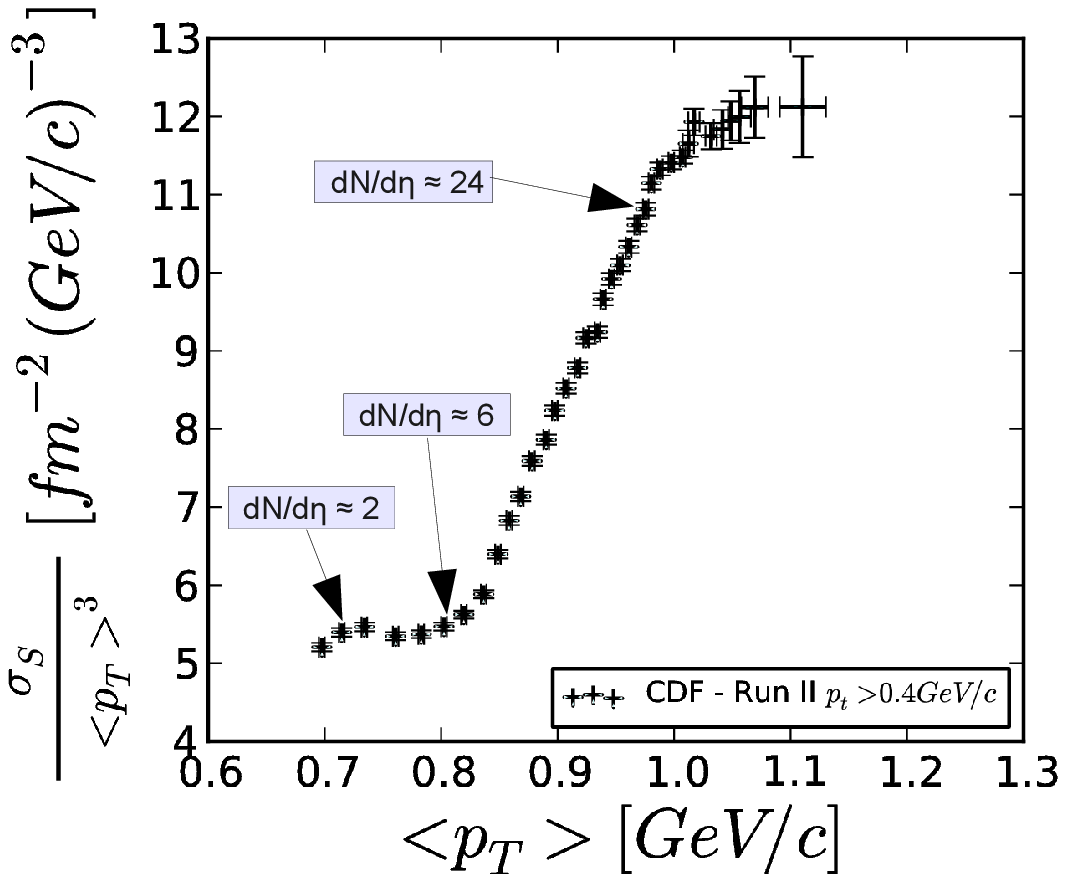}
	}\\
	\subfloat[]{
		\label{fig:sigmapt3_atlas_7_pt01}
	\includegraphics[width=0.35\textwidth]
{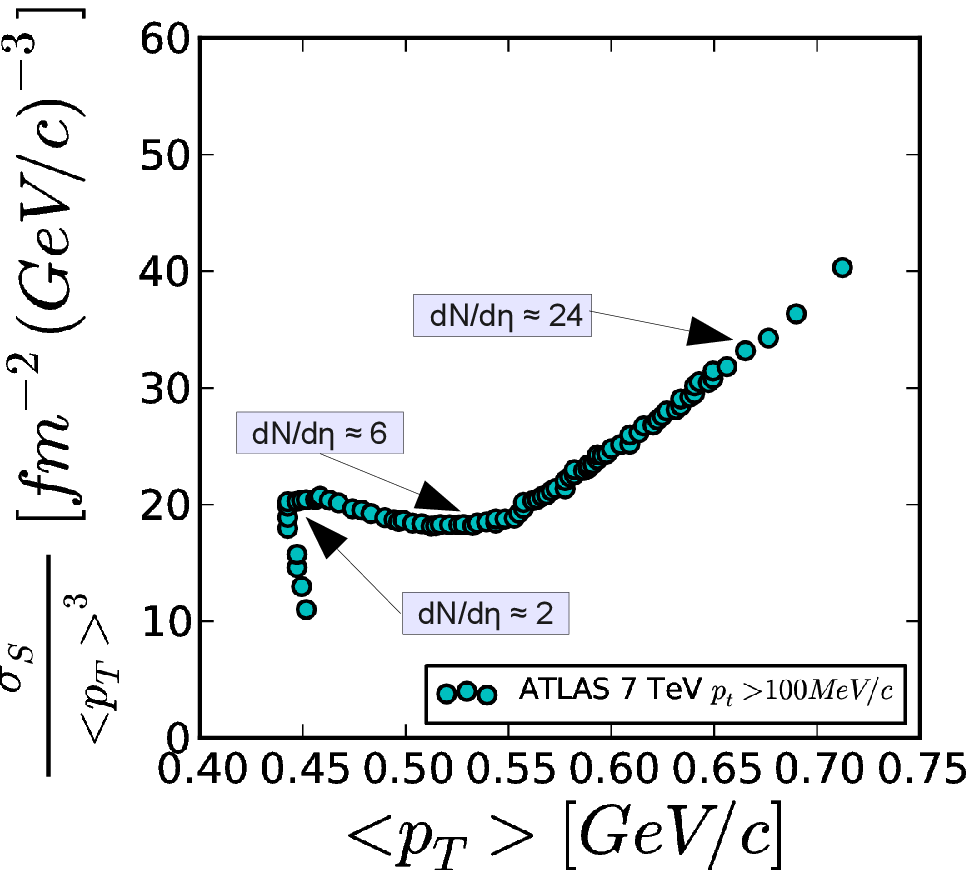}
	}\qquad    
	\subfloat[]{
		\label{fig:sigmapt3_cms_7_pt0}
	\includegraphics[width=0.35\textwidth]
{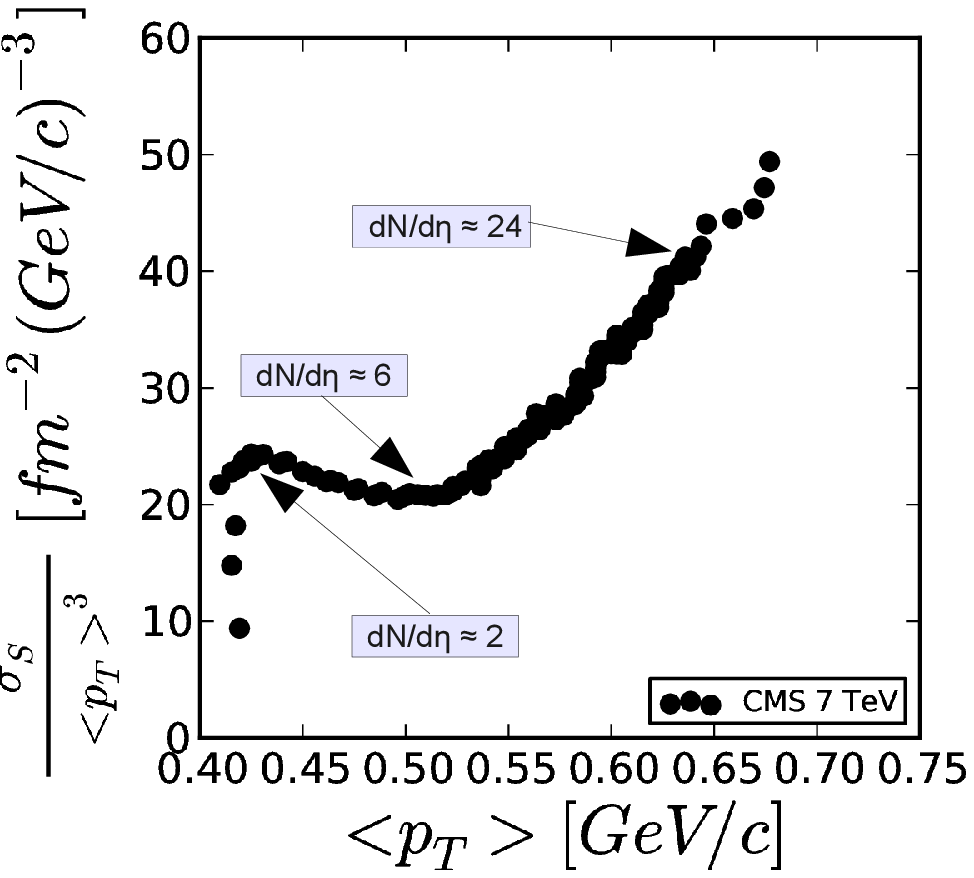}
	}
	
	\caption{$\sigmaptcube$~vs~$\meanpt$ plots. Area $S$ from ``$5^\mathrm{th}$
	degree polynomial + constant after $\dndeta > 7.5$'' fit.\\
	\ref{fig:sigmapt3_cdf_run1}: CDF Run I at $\sqrt{s}$ = 1.8 TeV, $\pt > 0.4$ GeV/c, $|\eta| < 1.0$,
	Minimum Bias.\\
	\ref{fig:sigmapt3_cdf_run2}: CDF Run II at $\sqrt{s}$ = 1.96 TeV, $\pt > 0.4$ GeV/c, $|\eta| < 1.0$,
	Minimum Bias + High multiplicity trigger.\\
	\ref{fig:sigmapt3_atlas_7_pt01}: ATLAS at $\sqrt{s}$ = 7 TeV, $\pt > 0.1$ GeV/c, $|\eta| < 2.5$,
	Minimum Bias.\\
	\ref{fig:sigmapt3_cms_7_pt0}: CMS at $\sqrt{s}$ = 7 TeV, $\pt > 0$ GeV/c, $|\eta| < 2.4$.
	Minimum Bias.
	}
	\label{fig:sigmaptcube_2}
	\end{center}
\end{figure}

Starting from $\sigma_S$ and $\meanpt$, we plotted
$\sigmaptcube$~vs~$\meanpt$ curves, as an
experimental approximation of $\sigma/T^3$~vs~$T$
curves. See Figs.~\ref{fig:sigmaptcube_1} and~\ref{fig:sigmaptcube_2}.
We obtained very similar $\sigmaptcube$~vs~$\meanpt$ curves from other pp and
p$\antip$ experiments~\cite{CDF_PTVSN_RUN1_Acosta2002,
E735_PTVSN_Alexopoulos1993, UA1_PTVSN_Albajar1990,
ISR_SFM_PTVSN_Breakstone1987} (not shown).

In figures, we put labels with corresponding $\dndeta$ values for interesting
points, in order to relate these points to the characteristic values in
$\dndeta$. In the different plots, different regions are recognizable. In
particular, in all plots we see that from the $\sigmaptcube$ value corresponding
to
$\dndeta \simeq 2$, up to a value correspondent to $\dndeta \simeq 6$, the curve
is almost
flat, then rises very quickly. This behavior is similar to the one in
$\sigmatcube$
curve, in presence of crossover, starting from a state of matter, identified by
$\sigmatcube$ nearly constant (region $2 \lesssim \dndeta \lesssim 6$),
and a crossover starting at $\dndeta \simeq 6$
(Fig.~\ref{fig:theoretical_lattice}).
Besides, in plots with many points at high
$\dndeta$ values (ATLAS with $\pt > 2500$ MeV/c, and CDF Run II 1960 GeV with
$\pt > 400$ MeV),
we observe a strong slope change around corresponding $\dndeta$ values of about
24 or higher.
It's worth noting that what seems to be a different behavior in the left side
for ATLAS with $\pt > 2500$ MeV/c (Fig.~\ref{fig:sigmapt3_atlas_7_pt2500}), is
only due to the fact that all points correspond to $\dndeta \gtrsim 7$,
apart from the first point, which correspond to $\dndeta \simeq 3.4$.
The ratio between $\sigmaptcube$ values corresponding to
$\dndeta \geq 24$ and
those corresponding to $\dndeta \leq 6$ varies from 2 to 3, depending on the
area
calculation method used for the estimation of $\sigma_S$. This ratio  in the
case of
EOS would correspond to the ratio between the number of the degrees of freedom
of the state  before and after the transition or the crossover. We note that for
small size
systems as it would be in the pp case, the jump in entropy density is
considerably reduced~\cite{Elze1986, Bazavov2007, Palhares2010} in
comparison to the
theoretical
infinite volume case. In plots with $\pt > 100$ MeV/c (ATLAS 7 TeV) or
$\pt > 0$ (CMS 7 TeV),
the first points have constant $\meanpt$ with varying $\dndeta$, which
leads to an initial steep rise. After that, the curves assume the same
behavior of previously seen plots.

\subsection{Sound velocity $\csquare$}
\label{subsec:methods_csquare}
\begin{figure}
	\begin{center}
	\subfloat[]{
		\label{fig:c2vseps_alice_09_pt05_vspt_doublefit}
\includegraphics[width=0.45\textwidth]
{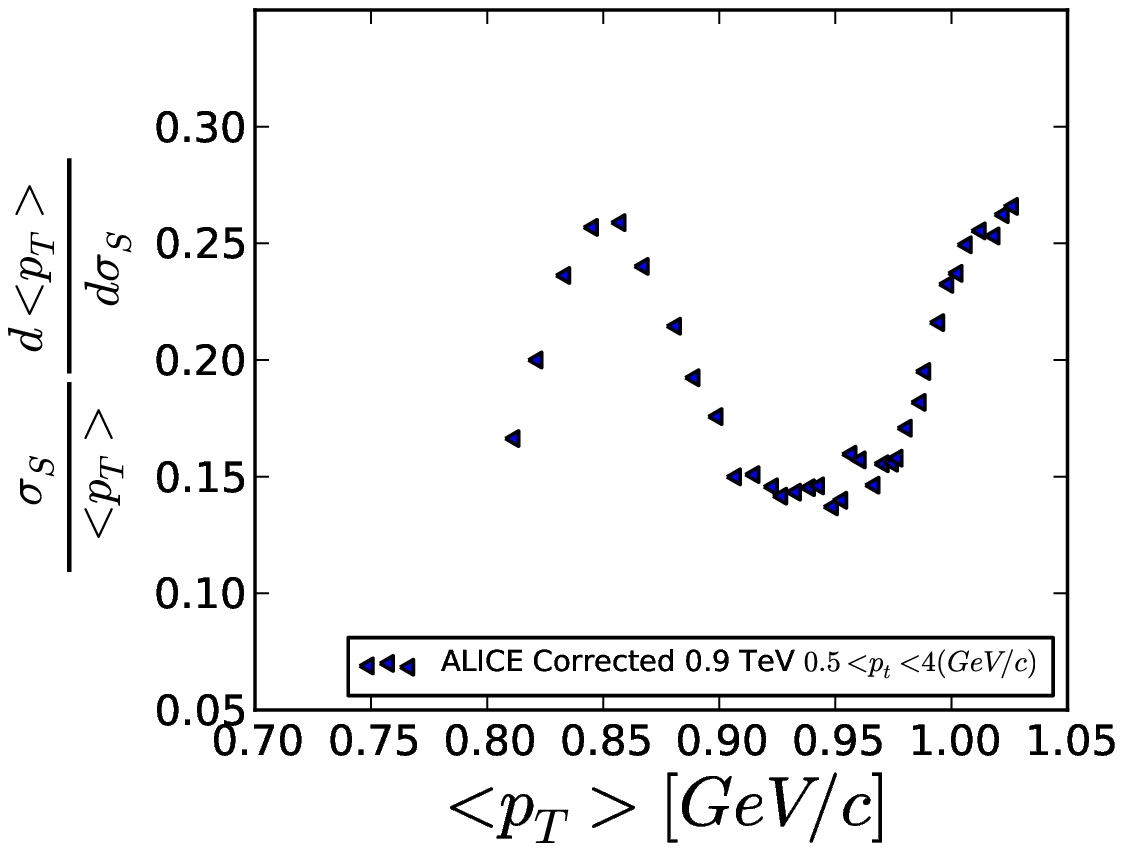}
	} 
\subfloat[]{
		\label{fig:c2vseps_alice_09_pt05_vseps_doublefit}
		\includegraphics[width=0.45\textwidth]
	{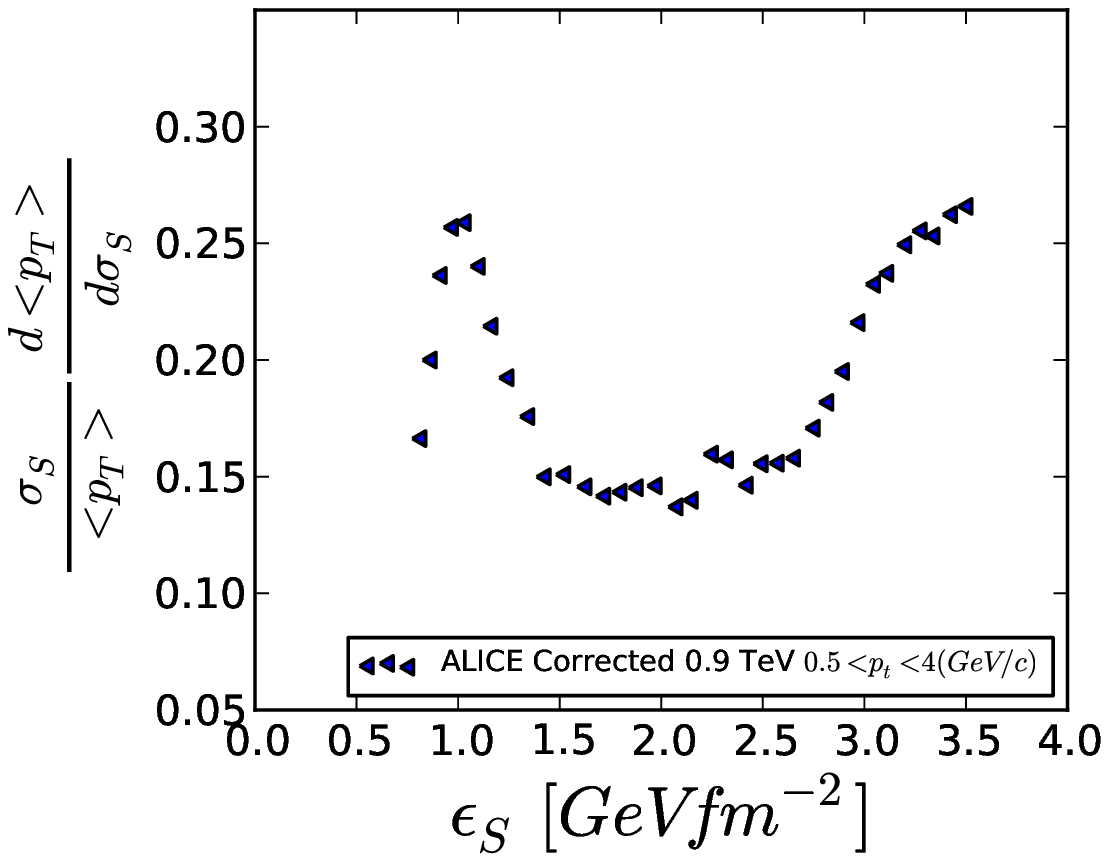}
	}\\
	\subfloat[]{
		\label{fig:c2vseps_cdf_run2_vspt_rconstantafter7_5}
	\includegraphics[width=0.45\textwidth]
{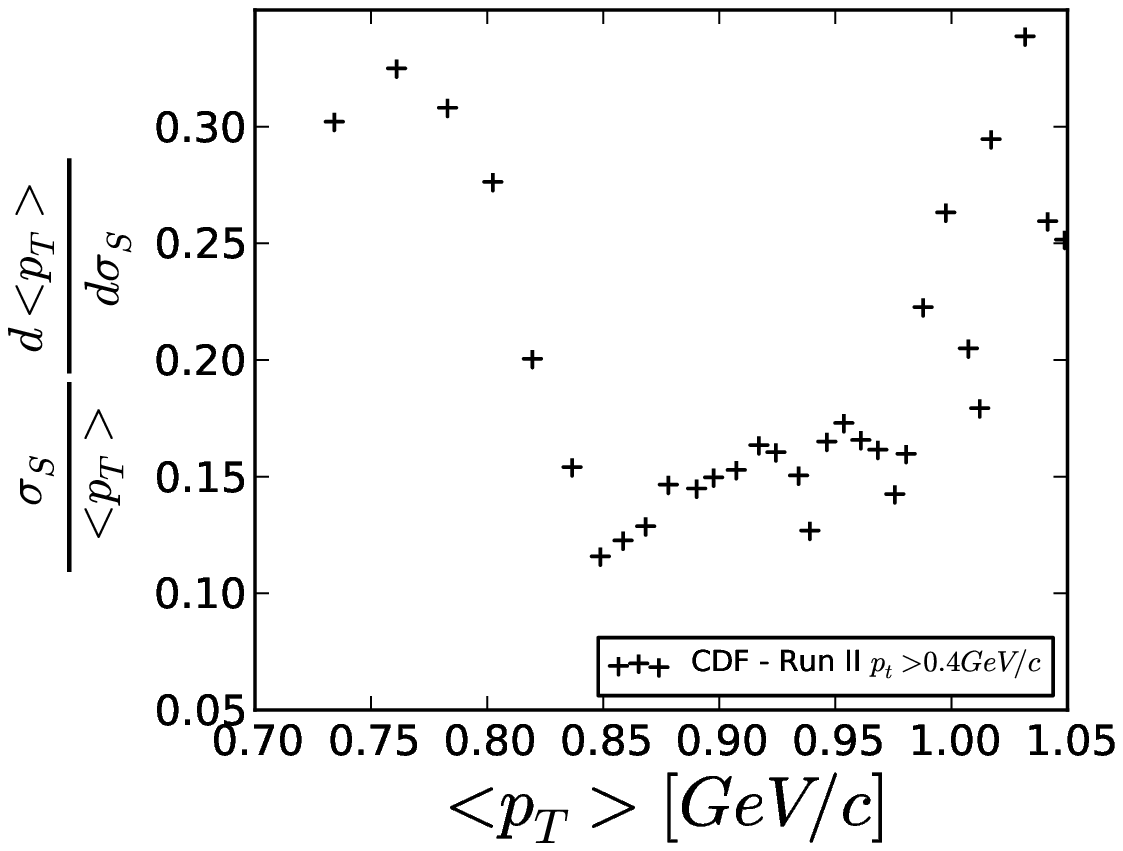}
	}   
	\subfloat[]{
		\label{fig:c2vseps_cdf_run2_vseps_rconstantafter7_5}
	\includegraphics[width=0.45\textwidth]
{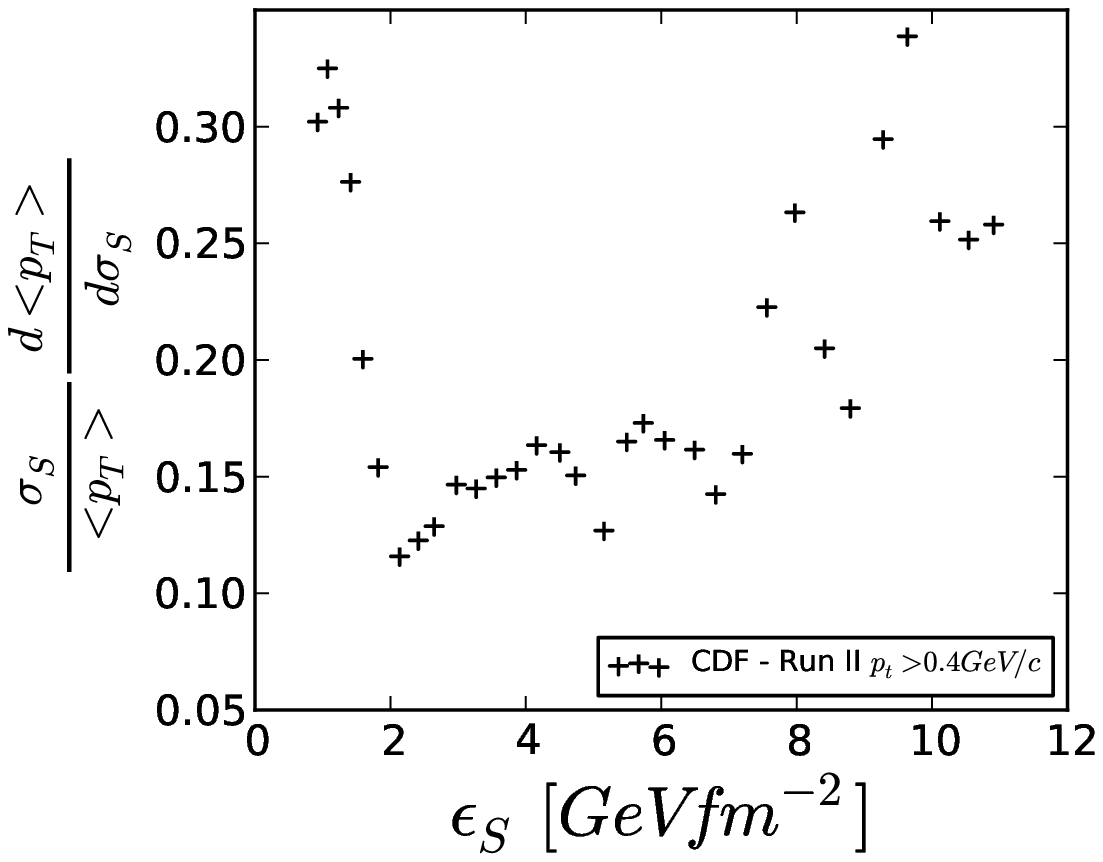}
	}
	
	\caption{$\csquare = \dfrac{\sigma_S}{\meanpt} \cdot \dfrac{d\meanpt}{d\sigma_S}$
	vs $\meanpt$ or $\epsilon_S$, using two different fits for area $S$ estimation.\\
	ALICE at $\sqrt{s}$ = 0.9 TeV, $0.5 < \pt < 4$ GeV/c, $|\eta| < 0.8$,
	Minimum Bias;
	area $S$ from ``$5^\mathrm{th}$ degree polynomial +
	linear in $\dndeta^{1/3}$ after $\dndeta > 7.5$'' fit.\\
	\ref{fig:c2vseps_alice_09_pt05_vspt_doublefit}: $\csquare$ vs $\meanpt$,
	\ref{fig:c2vseps_alice_09_pt05_vseps_doublefit}: $\csquare$ vs $\epsilon_S$.\\
	CDF Run II at $\sqrt{s}$ = 1.96 TeV, $\pt > 0.4$ GeV/c, $|\eta| < 1.0$,
	Minimum Bias + High multiplicity trigger;
	area $S$ from ``$5^\mathrm{th}$ degree polynomial +
	constant after $\dndeta > 7.5$'' fit.\\
	\ref{fig:c2vseps_cdf_run2_vspt_rconstantafter7_5}: $\csquare$ vs $\meanpt$,
	\ref{fig:c2vseps_cdf_run2_vseps_rconstantafter7_5}: $\csquare$ vs $\epsilon_S$.}
	\label{fig:c2vseps}
	\end{center}
\end{figure}

One of the physical quantities used to characterize the state of a system is its
squared sound velocity, defined as $\csquare = \dfrac{\sigma}{T} \cdot
\dfrac{dT}{d\sigma}$, for constant V~\cite{Florkowski_BOOK_2010}.
In our study, we approximate it with $\csquare = \dfrac{\sigma_S}{\meanpt} \cdot
\dfrac{d\meanpt}{d\sigma_S}$. It is really
interesting that if $\meanpt$ is proportional to $T$ and if $\sigma_S$ is
proportional to the
entropy density, then the $\csquare$ value obtained in this approximation is
equal to the right value of $\csquare = \dfrac{\sigma}{T} \cdot
\dfrac{dT}{d\sigma}$, because proportionality constants cancel out. In
order to obtain
our $\csquare$ estimation, from $\meanpt$~vs~$\dndeta$ curves and from
$\sigma_S$ values, we
compute the
curve $\meanpt$~vs~$\sigma_S$, to which we apply numerical derivation. We cope
with the
statistical fluctuation in data points using a combination of
Gaussian and Savitzky-Golay filters~\cite{Savitzky1964}.
Examples of $\csquare$~vs~$\meanpt$ curves are shown in 
Figs.~\ref{fig:c2vseps_alice_09_pt05_vspt_doublefit}
and~\ref{fig:c2vseps_cdf_run2_vspt_rconstantafter7_5}.

The so obtained $\csquare$ estimation resembles the typical shape of a
phase
transition or a crossover: a descent, a minimum region and a following rise,
as it's also obtained
analytically from EOSs which present a phase transition or a crossover.
The minimum value reached by the estimation of $\csquare$ in the different
experimental curves varies from 0.08 to 0.18 and could correspond to what it's
called the EOS softest point~\cite{Shuryak_Softest_Point_1986}.

Recently
Refs.~\cite{Castorina_Cleyman_Satz_2010,Srivastava_PercolationandDeconfinement,
Chojnacki_Florkowski_2007a} estimate $\csquare$ minimum value for realistic
EOS to be around 0.14.
From $\epsilon_S \simeq \meanpt \cdot \sigma_S$ we compute
$\csquare$~vs~$\epsilon_S$ curves, that are
approximations of $\csquare$ vs energy density. We report these curves in
Figs.~\ref{fig:c2vseps_alice_09_pt05_vseps_doublefit}
and~\ref{fig:c2vseps_cdf_run2_vseps_rconstantafter7_5}.

In this case, $\epsilon_S$ values are calculated using $\meanpt$ values
from
$\meanpt$~vs~$\dndeta$ curves with no $\ptmin$ cut at corresponding energies,
estimated in correspondence with the different $\dndeta$ values. As
Figs.~\ref{fig:c2vseps}  show,
the numerical estimation of $\csquare$ vs $\epsilon_S$,
is characterized by a maximum at low energy density followed by a minimum
region, which is obtained for $\epsilon_S$ values in range $1.5-2.0$ GeV/fm$^2$,
and a subsequent rise.

We note that $\epsilon_S$ as computed here is an estimation of the energy
density for pseudorapidity unit and unit of transverse area. In order to
estimate the volume energy density this should be divided by ct.

\section{Discussion}
\label{sec:discussion}
The shape of the $\sigmaptcube$ approximation to the EOS is very similar,
using both $\rinv$
from the fit on all $\dndeta$ space and $\rinv$ fitted up to $\dndeta$ = 7.5 and
then
maintained constant. It slightly varies when using the area from the impact
parameter model, but the slope change at $\sigmaptcube$ values corresponding to
$\dndeta$ around 6 is still present, as well as the change at $\sigmaptcube$
values corresponding to $\dndeta$ about 24.

In order to avoid possible systematics due to calculation involved in the
area definition, we plotted directly
$\dndeta/\meanpt^3$ vs $\meanpt$: this is equivalent to obtain
$\sigmaptcube$ curves considering a
transverse section which is constant for all $\dndeta$ values.
For space reason we don't show these plots in this paper.
In this case the shape doesn't resemble an EOS shape anymore, but the slope
changes at $\dndeta \simeq 6$ and $\dndeta \simeq 24$ are still present, because
they are contained in the $\meanpt$~vs~$\dndeta$ correlation.

The shape of the curves obtained from experimental data
($\sigmaptcube$~vs~$\meanpt$, $\csquare$~vs~$\meanpt$ and $\csquare$ vs energy)
depends on experimental $\meanpt$~vs~$\dndeta$ curves and from the value
of the area used to obtain density sigmas. Systematic errors in $\meanpt$,
$\dndeta$, and $\rinv$
measurements don't lead to appreciable variations in $\meanpt$ vs $\sigma_S$
behavior, which is what we are interested on.

It seems to us that the main result of this work is that putting together
experimental data of $\meanpt$ vs $\dndeta$ and $\rinv$ vs $\dndeta$, curves are
obtained which are the reproduction of theoretical EOS curves.

\begin{figure}
	\begin{center}
	\subfloat[]{
		\label{fig:MC_atlas_09_pt05}
\includegraphics[width=0.3\textwidth]
{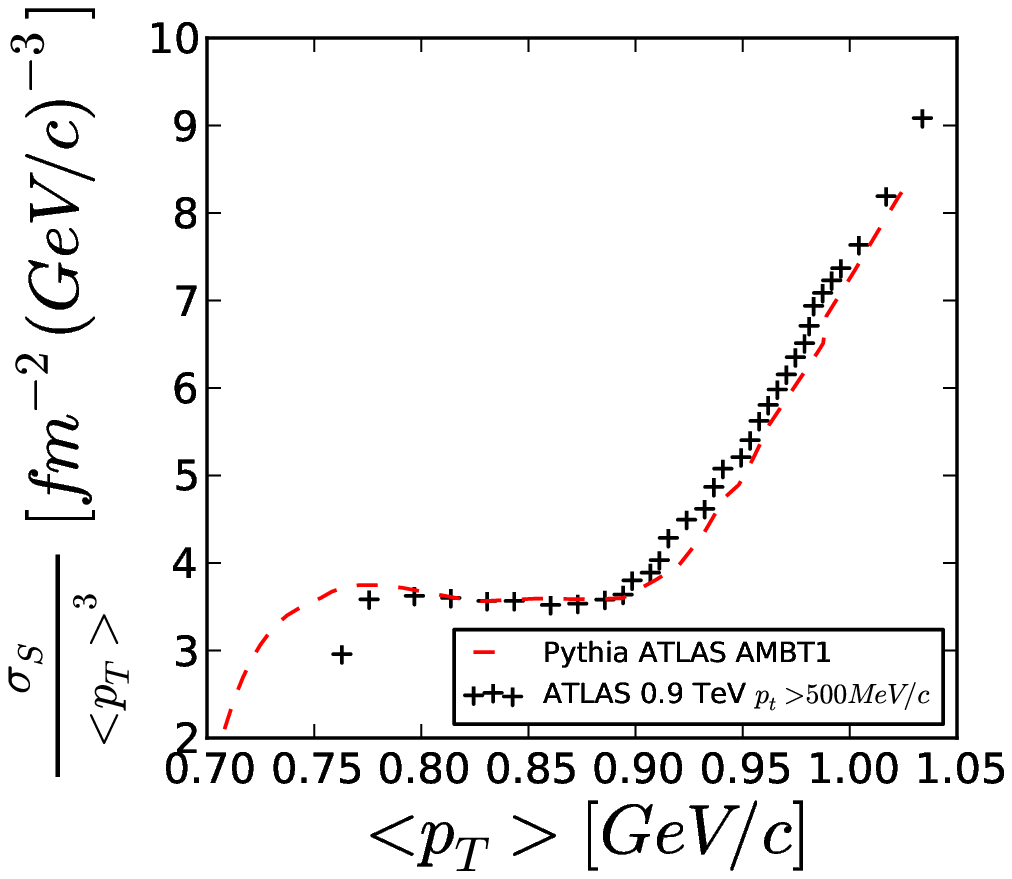}
	} 
\subfloat[]{
		\label{fig:MC_atlas_7_pt05}
		\includegraphics[width=0.3\textwidth]
		{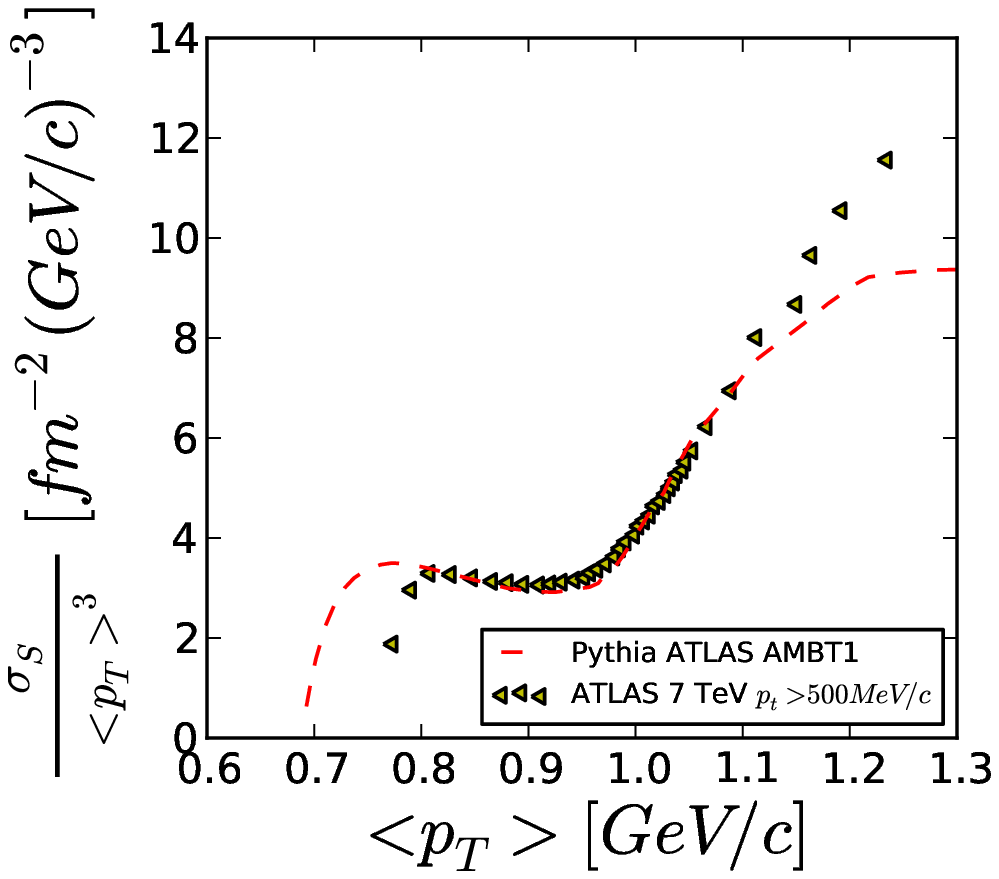}
	}
	\subfloat[]{
		\label{fig:MC_atlas_7_pt01}
	\includegraphics[width=0.23\textwidth]
{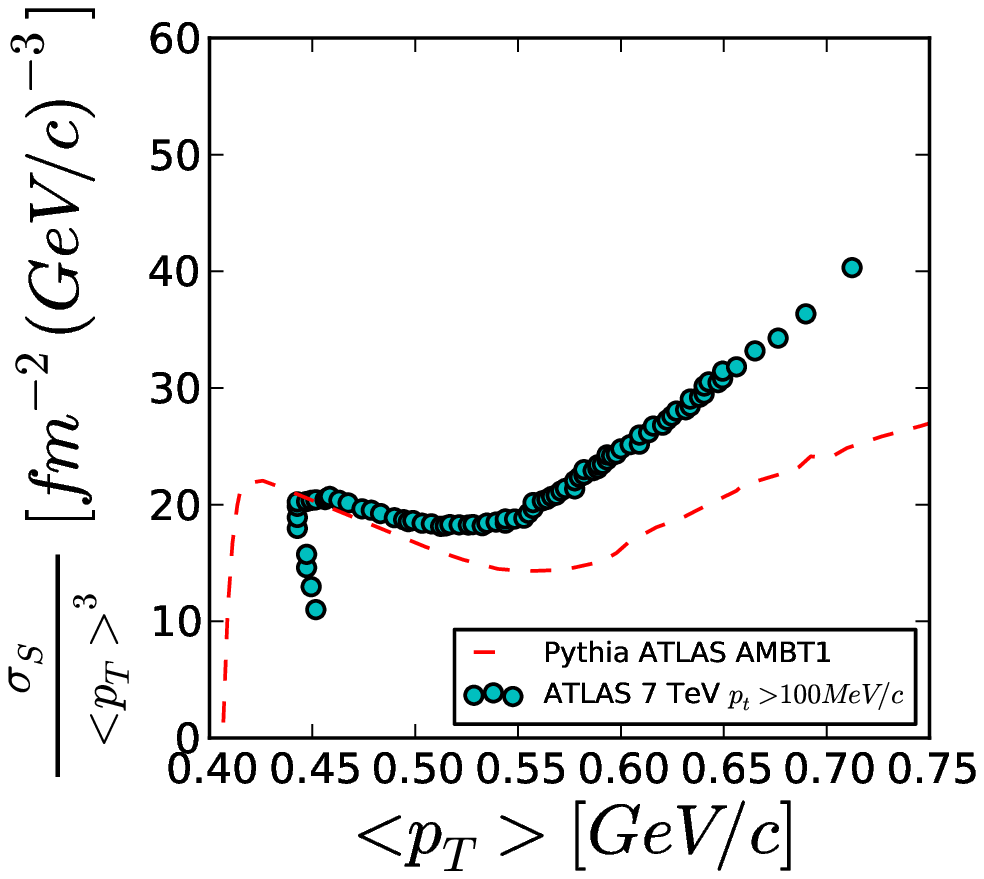}
	}\\
	\subfloat[]{
		\label{fig:MC_CMS_7_pt01}
	\includegraphics[width=0.23\textwidth]
{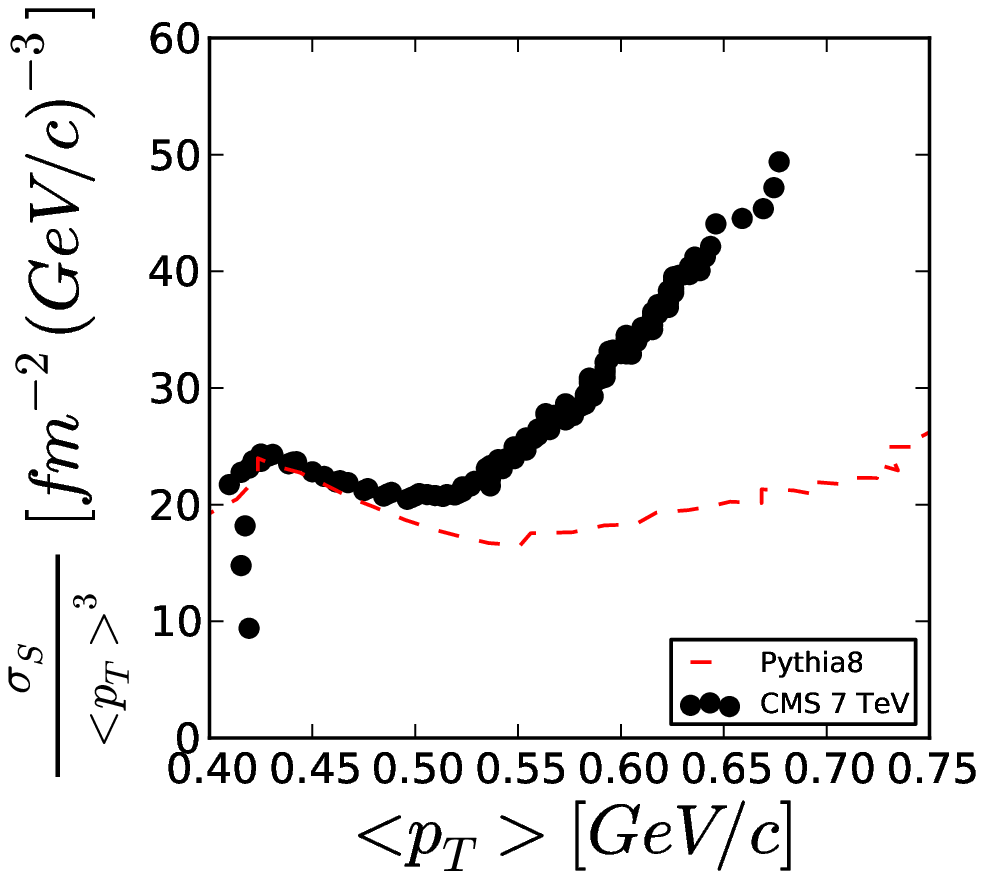}
	}
	\subfloat[]{
		\label{fig:MC_atlas_7_pt25}
	\includegraphics[width=0.3\textwidth]
{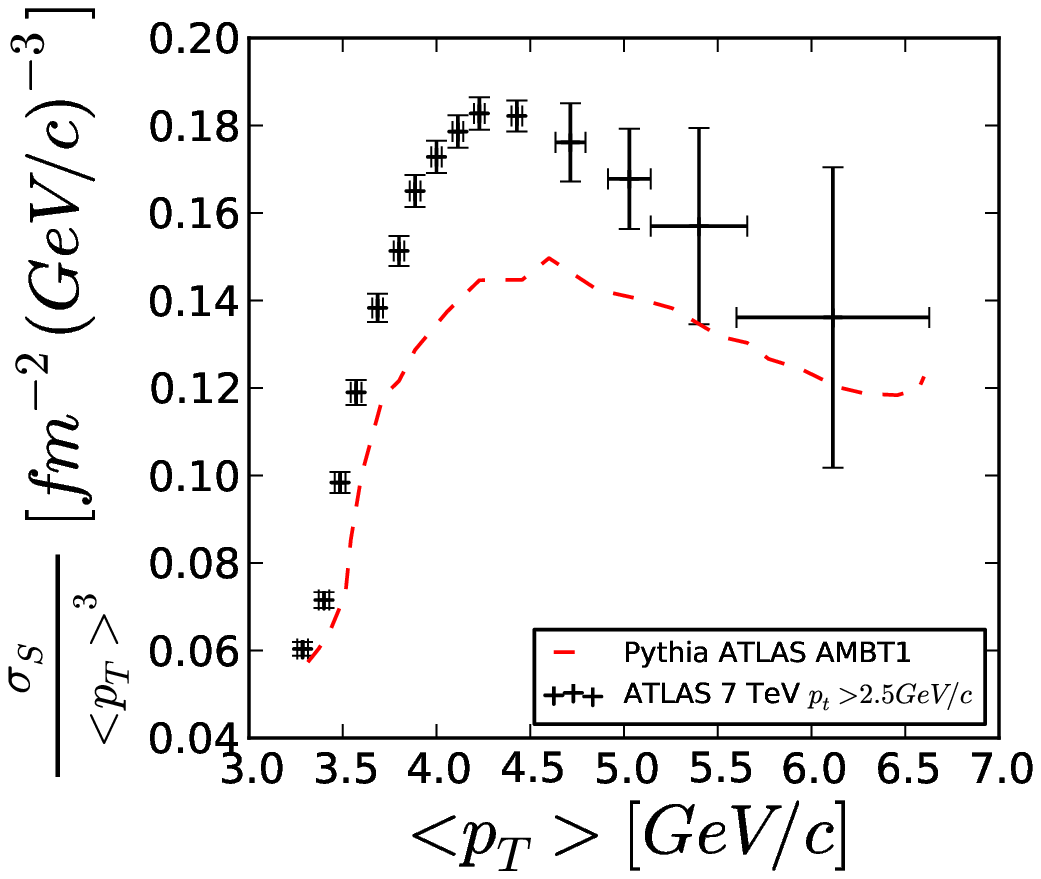}
	}   
	\caption{$\sigmaptcube$~vs~$\meanpt$. Comparison with models.\\
	\ref{fig:MC_atlas_09_pt05}: ATLAS at $\sqrt{s}$ = 0.9 TeV, $\pt > 0.5$ GeV/c, $|\eta| < 2.5$,
	Minimum Bias
	and \emph{Pythia ATLAS AMBT1}.\\
	\ref{fig:MC_atlas_7_pt05}: ATLAS at $\sqrt{s}$ = 7 TeV, $\pt > 0.5$ GeV/c, $|\eta| < 2.5$,
	Minimum Bias
	and \emph{Pythia ATLAS AMBT1}.\\
	\ref{fig:MC_atlas_7_pt01}: ATLAS at $\sqrt{s}$ = 7 TeV, $\pt > 0.1$ GeV/c, $|\eta| < 2.5$,
	Minimum Bias
	and \emph{Pythia ATLAS AMBT1}.\\
	\ref{fig:MC_CMS_7_pt01}: CMS at $\sqrt{s}$ = 7 TeV, $\pt > 0$ GeV/c, $|\eta| < 2.4$,
	Minimum Bias
	and \emph{Pythia8}.\\
	\ref{fig:MC_atlas_7_pt25}: ATLAS at $\sqrt{s}$ = 7 TeV, $\pt > 2.5$ GeV/c, $|\eta| < 2.5$,
	Minimum Bias
	and \emph{Pythia ATLAS AMBT1}. }
	\label{fig:MC}
	\end{center}
\end{figure}

Regarding model comparison, we obtained $\sigmaptcube$~vs~$\meanpt$ plots
starting from Montecarlo curves (\emph{Pythia ATLAS AMBT1} and \emph{Pythia8}
for ATLAS and CMS experiment respectively), which are shown in Fig.~\ref{fig:MC}.

Some models on which tuning has been done, for
example with CDF Run II data at 1960 GeV for $\pt >$ 400 MeV/c, well reproduce
the $\meanpt$~vs~$\dndeta$ curve at higher (7 TeV) or lower (0.9 TeV) energies
with $\pt > 500 $MeV/c. It is clear that in these cases, starting from the
$\meanpt$~vs~$\dndeta$ curves of models and
using Bose Einstein correlation or the impact parameter--multiplicity relation
for $\sigma_S$ estimation, curves similar to the experimental ones are obtained.
On the other hand, models don't predict well $\meanpt$~vs~$\dndeta$ curves with
low $\ptmin$, and consequently $\sigmaptcube$~vs~$\meanpt$ curves as shown for
the comparison of models at 7 TeV for CMS and ATLAS data, respectively with
$\pt > 0$ MeV/c and $\pt > 100$
MeV/c~\cite{CMS_PTVSN_Khachatryan2011,ATLAS_PTVSN_1_Aad2010,
ATLAS_PTVSN_2_Aad2011,CMS_PTVSN_2010}.

The interpretation of curve shapes as experimental
``estimation'' of EOS depends on how much likely are the correspondences
between $\meanpt$ and $T$, and between measured $\sigma_S$ and entropy.

\section{Conclusion}
\label{sec:conclusion}

The result we consider to be the most important is the following: in many
experiments~\cite{
ALICE_PTVSN_Aamodt2010,CMS_PTVSN_Khachatryan2011,ATLAS_PTVSN_1_Aad2010,
ATLAS_PTVSN_2_Aad2011,
CDF_PTVSN_RUN1_Acosta2002,CDF_PTVSN_RUN2_Aaltonen2009a,
E735_PTVSN_Alexopoulos1993, UA1_PTVSN_Albajar1990, ISR_SFM_PTVSN_Breakstone1987}
from 31 GeV to 7000 GeV,
starting from $\meanpt$ vs $\dndeta$ and using results from measures of radii
with Bose
Einstein correlation or from a model that relates impact parameter and
multiplicity, we obtained that $\sigmaptcube$~vs~$\meanpt$ and 
$\csquare = \dfrac{\sigma_S}{\meanpt} \cdot
\dfrac{d\meanpt}{d\sigma_S}$ reproduce
the shape of hadronic matter EOSs and squared sound velocity respectively,
in presence of crossover or phase transition. From the
plots, a neat change around $\dndeta$ around 6, where the crossover or the
phase transition seems to start, and another possible change at
$\dndeta$ around 24 are observed. The curve $\csquare$~vs~$\epsilon_S$ has a minimum
around a ``transverse'' energy density of about 1.5 GeV/fm$^2$.

In order to
understand if these behaviors have a real physical meaning or are just casual,
results of measures in the following regions should be compared: 2 $ \lesssim
\dndeta
\lesssim $ 6, $\dndeta \gtrsim$ 6, 6 $\lesssim \dndeta \lesssim$ 24 and $\dndeta
\gtrsim $ 24.





\section*{References}
\label{sec:references}

\bibliographystyle{model1-num-names}
\bibliography{eos}







\end{document}